# Determining Secondary Attributes for Credit Evaluation in P2P Lending

Revathi Bhuvaneswari
*Department of CIS*
*Fordham University*
New York, USA
rbhuvaneswari@fordham.edu

Antonio Segalini
*Department of CIS*
*Fordham University*
New York, USA
asegalini@fordham.edu

*Abstract*—There has been an increased need for secondary means of credit evaluation by both traditional banking organizations as well as peer-to-peer lending entities. This is especially important in the present technological era where sticking with strict primary credit histories doesn't help distinguish between a 'good' and a 'bad' borrower, and ends up hurting both the individual borrower as well as the investor as a whole. We utilized machine learning classification and clustering algorithms to accurately predict a borrower's creditworthiness while identifying specific secondary attributes that contribute to this score. While extensive research has been done in predicting when a loan would be fully paid, the area of feature selection for lending is relatively new. We achieved 65% F1 and 73% AUC on the LendingClub data while identifying key secondary attributes.

## I. Problem Statement

An exceedingly interconnected world that relies on technology requires secondary methods of identity checks as well as credit histories. People who apply to receive credit cards or mortgages with minimal or no credit history are harmed, including those from other countries. Therefore, a secondary method of creditworthiness needs to be determined that can look beyond a strict credit score.

## II. Related Work

Authors of [3] have expressed in their research that class imbalance between good borrowers and bad borrowers has a significant impact on the accuracy of the prediction made by their model. Since it is common to observe a greater percentage of good loans than the bad loans in studies of P2P credit, by default the classifiers tend to be biased towards the majority class (in this case the good borrowers), which thereby affects the classifiers prediction accuracy. To tackle this class imbalance, they proposed utilizing class rebalancing techniques such as Under Sampling, Over Sampling, Under & Over Sampling and Random Over-Sampling Examples (ROSE) to obtain a rebalanced sample which is almost the same size as that of the original sample. They also proposed deploying multiple probability prediction algorithms such as Generalised Additive Model (GAM), Naive Bayes (NB), Random Forest (RF) and Extreme Gradient Boosting (XGBoost) and combined them using regularized logistic regression to improve the prediction accuracy as well as prevent overfitting of the model.

Authors of [1] focus on the potential of improving the existing credit models and loan screening techniques by deploying Deep Neural Networks (DNN) along with Logistic Regression (LR). The model divided the dataset into two phases: loan rejection prediction in the first phase and default risk for approved loans in the second phase. They concluded that by appropriate feature selections while cleaning the data, and by deploying LR on the first phase (rejection recall ≈ 85%) and the DNN on the second phase (default recall ≈ 75%), significant improvement in the prediction accuracy can be achieved in an automated way. They were able to demonstrate the current discrepancies in loan screening + default prediction by deploying their model on loans for small businesses.

Authors of [7] have proposed developing a Decision Support System (DSS) that goes beyond the traditional P2P lending credit scoring system and focuses upon lender profitability by involving the Internal Rate of Return (IRR) as a profit scoring measure. They are able to justify this DSS by concluding that the variables for predicting loan default are different than that of loan profitability. From their study, the authors concluded the P2P loan market is not completely efficient, and that borrowers who have a high probability of defaulting can also be profitable. Therefore they emphasize that there exists a nonlinear relationship between the variables, and use of nonlinear techniques such as Decision Trees allows for this nonlinear relation between IRR and predictive factors. Such a DSS can aid in improving this imperfect P2P lending market, and move towards resembling it to a perfect market.

In [10] "How the machine 'thinks': Understanding opacity in machine learning algorithms," the authors discuss the difficulty opacity and lack of knowledge can have in machine learning, touching on the difficulty of accurately predicting credit score with limited information. This paper finds that recognizing the particular opacity in something like credit scoring will help determine how to make algorithm choices that minimize that particular opacity.

Authors of [11] investigated the determinants of default in P2P loans. These three authors looked at different subsets of attributes to see the default ratios on loans. While not perfectly a match for predicting grade, looking at defaulting and reverse engineering the results can also allow for a commonality.

### III. Dataset Description

The dataset used for this project is taken from Kaggle[5] and contains information on loan borrowers in the P2P lending platform called LendingClub. The dataset consists of two separate CSV files - accepted loans and rejected loans. Each data point in these files represents a unique borrower and is distinguishable by their respective IDs. The time period of all loans is between 2007 and 2018.

The accepted loans file contains information on 2,260,701 borrowers arranged in 151 different columns. These columns contain basic loan information like loan amount and interest as well as detailed background information on the borrower like annual income, employment title, debt-to-income ratio, housing type, and so on. There are few columns dedicated to a borrower's rating such as their FICO score and the grade assigned by LendingClub based on their credit risk. This grade column, which is categorical in nature with values in [A, G], will be used as our target column for this project. The majority of the columns are of type float, with the rest falling under the type of string or int.

The rejected loans file, on the other hand, contains information on 27,648,141 borrowers arranged in 9 different columns. These columns contain basic information such as loan amount requested, debt-to-income ratio, state, employment length, etc. which are also present in the accepted loans file. A new column called the risk score is available instead of the LendingClub assigned letter grade for the borrower. This score is of type float with an average of around 600, which can help signify why the loan was rejected. This file will be further analyzed in our project to get a better understanding of the attributes that make a 'good' borrower.

### IV. Approach

The project looks to find characteristics that will best project whether a borrower would be approved or rejected for a loan, and the risk factor associated with them. The original dataset from LendingClub had 151 initial attributes. The first set of attribute decisions were made by looking at each individual attribute as well as the data dictionary that accompanied the dataset to look for minimum viability. Attributes that were irrelevant were removed leading to 72 attributes and one target remaining, which is the grade.

Fig. 1. Sample of Initial Attributes

#### A. Preprocessing

From there, categorical attributes were broken down using hashing and encoding techniques. For example, the state attribute was split into 8 distinct columns in order to allow for data manipulation. After the ordinal, one-hot, and hashing encoding was performed on the categorical attributes, there were 93 attributes and one target remaining.

In order to consolidate attributes, a cross-correlation analysis was done to see if any two attributes were highly correlated (1) with the target grade, and (2) with other attributes in the dataset. A subset of this analysis is shown in the heat map in Fig. 2.

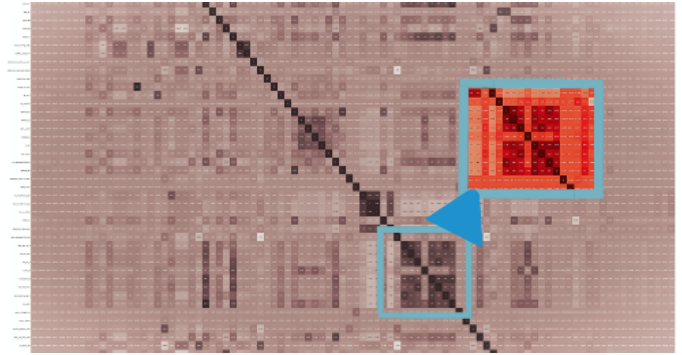

Fig. 2. Heat Map of Attributes

Running across all 93 attributes, the heat map showed areas of high correlation with grade but also a high correlation between attributes, leading to potential areas of consolidation. For example, the last_fico_range_high attribute was highly correlated with the last_fico_range_low attribute, and multiple attributes related to the Fico Score were highly correlated with one another, shown in the pop-out portion in Fig. 2.

During this initial attribute selection, those areas of cross-correlation were investigated when the absolute value of correlation between the two attributes was ≥ 0.9. Fig. 3 shows a sample of initial attributes as well as instances of correlation and actions taken due to those attributes.

Fig. 3. Attribute Descriptions and Samples

As seen in Fig. 3, highly correlated items, such as those outlining the fico range, were consolidated. Items that logically would not make sense as predictive measurements were also dropped. This included int_rate, as the interest rate for a loan was determined by a grade and logically could not exist as a predictive measure of a grade.

Each individual attribute was also investigated using the possible values as well as the data dictionary when looking at missing values. For example, annual_inc was filled with the minimum if it was missing. As seen in Fig. 3, the annual_inc attribute shows income values. Therefore, a missing value would be treated as having "no income" or 0.0.

Other instances such as the mo_sin_old_rev_tl_op were filled with the maximum, as someone who had no value would have had no recent openings of new accounts, and therefore would be the maximum distance between the day of the loan and the day of opening a new account.

By using intentionality and taking the time in advance of running analysis, a solid foundation was made for research, which impacted the viability of results in a positive manner. After consolidating, 84 attributes remained with target grade. Finally, once attributes' missing values were filled and taken care of, a final correlation analysis was run between the 84 remaining attributes and the target grade. Fig. 4 shows the correlation analysis of the remaining attributes.

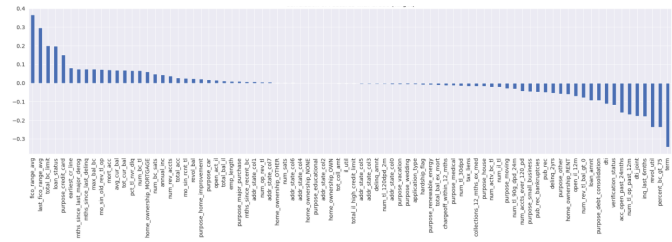

Fig. 4.  Correlation of Features with Grade (Target)

The investigation looked at highly positively-correlated items as well as highly negatively-correlated attributes, as a high negative correlation could also be predictive. This correlation analysis showed that attributes such as total bank card limits as well as specific purposes were highly correlated with grade and items like the term of the loan and other specific purposes were negatively correlated.

To include both items and a solid subset of attributes to run our tools against, we included all attributes with an absolute value of 0.01 or greater. This led to 57 attributes and the target, grade, which became our dataset.

### B. Performance Measures

Traditional classification metrics such as Precision, Recall, and F1-Score will be used on the dataset to further build baselines and choose the best-performing model. Additionally, Area Under the Curve (AUC) obtained from the Receiver Operating Characteristic (ROC) curve will be used to measure the class-distinguishing capability of our models.

Since there is a class imbalance in the grade column, the per-class accuracy will also be determined, which is especially useful for our multi-class classification problem. Our aim is to find the best combination of secondary attributes for credit evaluation that gives the highest possible value for the metrics.

### C. Target Attribute

In the initial dataset, grade was given on an A to G scale, with A being the highest grade and G being the lowest grade. The breakdown heavily weighted higher grades, A, B, and C, and had significantly fewer lower grades of D through F, as shown in Fig. 5, which gives the initial breakdown of grades.

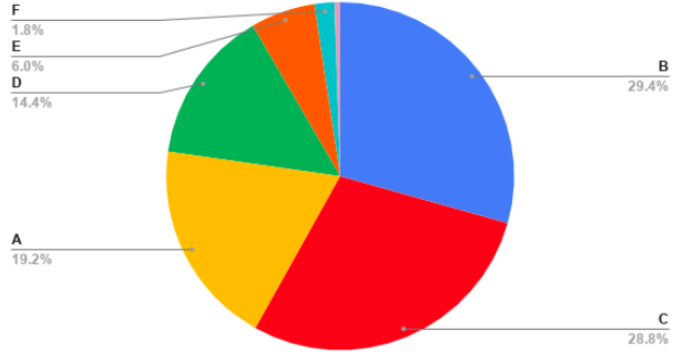

Fig. 5.  Initial Breakdown of Grade Averages (A to G)

In order to maintain a multiclass scenario but clean up the breakdown of grades, the grades were reclassified into "High risk", "Medium risk" and "Low risk". "High risk" included what were formerly grades D through G. "Medium risk" included grade C. "Low risk" included grades A and B. This led to a new breakdown shown as shown in Fig. 6.

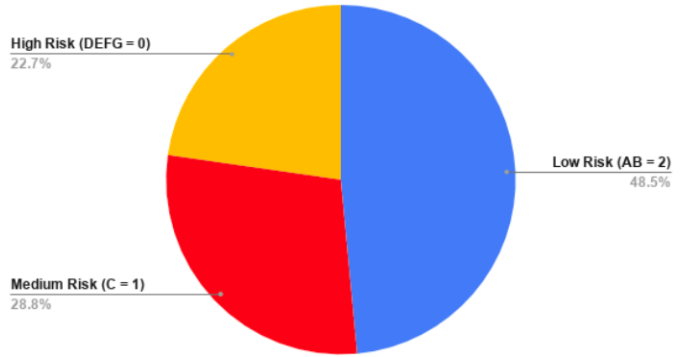

Fig. 6.  Updated Breakdown of Grade Averages (2, 1, 0)

This consolidation allowed for a cleaner vision of what users could potentially be grouped into from a business perspective. Furthermore, it allowed for better representation of different grades that would need oversampling or undersampling in order for a cleaner distribution.

### D. Methodologies

Given the distribution of grade, multiple sampling and scaling methods were taken into account. Firstly, four different classification algorithms were used in order to get a cross-swath of results from the given methods. Traditional classification algorithms such as Logistic Regression and SVM were used along with ensemble approaches of Bagging via Random Forest.

Another potential approach is to see if there are similarly-related groupings of characteristics that, together, can be more predictive than individual characteristics.

Therefore, we utilized the K-Means clustering algorithm as well to see if there is potential grouping in the pared-down dataset. Through a combination of these algorithms along with basic visualization and background knowledge, we can hopefully get to a point where we are identifying one or more characteristics that can assist in predicting whether or not a characteristic or group of characteristics can serve as a secondary creditworthiness check. Due to a large number of data points available, the processing time of our models will be taken into consideration for efficiency. Fig. 7 gives a general outline of decisions made.

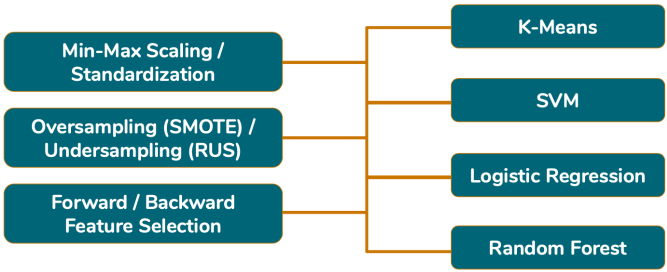

Fig. 7. Algorithms and Methods Used

Along with the multiple different algorithms, we also tested different standardization options, sampling, and feature selection. For standardization options, we looked into whether or not different scales would impact the resulting methods. This was also important when looking at K-Means to ensure that clusters were not over-influenced by the larger-scale differences for certain attributes.

Given the class imbalance issue, sampling methods such as oversampling using SMOTE and undersampling using RUS were looked at as well. Finally, especially when looking at Logistic Regression and Random Forest, feature selection could play an undeserved influence over results. In order to minimize the impact, we looked at ways to randomize the entire feature selection process.

The methodologies chosen were done in order to try to receive consistent results that could be relied upon and to eliminate the possibility of overtraining on the training sample. This was a way to deliver results that were reliable. The benefits to these methods are noted later where RUS helps eliminate large overfitting of data in Random Forest.

We broke down the initial dataset into 80% training and 20% testing. For the training portion, we used 5-fold cross validation in order to avoid the influence of outliers. This allowed for more consistent results across each method. The testing split was used later on with the models to simulate the effects of making a prediction on an unseen real-world dataset.

## V. EXPERIMENTS AND RESULTS

All four algorithms were initially run without sampling and scaling methods to build a baseline result that could show the effectiveness of each. The below evaluations and results will begin with the baseline models for all results and then continue to describe the initial results for each modeling technique. Given that this is a multiclass issue with a number of different attributes, the focus was on F1 and Area Under the Curve (AUC) scores when evaluating each model.

### A. Baseline Models

The four models were run through an initial baseline with no scaling or sampling methods. As mentioned in the *Methodologies* section, the baseline results were obtained using 5-fold validation on the 80% training split. Fig. 8 shows the initial F1 score across the models.

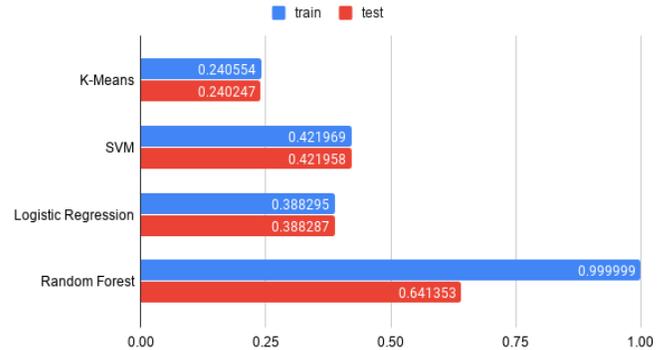

Fig. 8. Initial F1 Score of Baseline Models

As we can see looking at the F1 score, the training and testing results are similar for all except for Random Forest, which has significant overfitting. This result also appears when we look at the baseline AUC scores in Fig. 9.

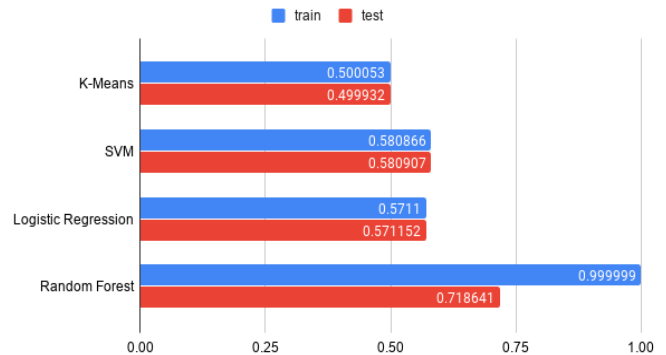

Fig. 9. Initial AUC Score of Baseline Models

Again, there is a close relationship between training and testing for all except for Random Forest. As mentioned previously, the decision to have different sampling and scaling methods is important here because it allowed for us to account for potential overfitting and produce a realistic model.

It should also be noted that K-Means produced poor baseline results compared to other models. Given the number of attributes and large dataset, this is in line with expectations due to the difficulty of working in such a large space in N-dimensions.

### B. K-Means

When setting up K-Means for modeling on the dataset, it was first important to find the correct number of clusters that should be used. This was done using the sum of WCSS or the weighted sum of squares and using the Elbow Method to determine where the WCSS flattened out. The WCSS was initially run without scaling, as shown in Fig. 10.

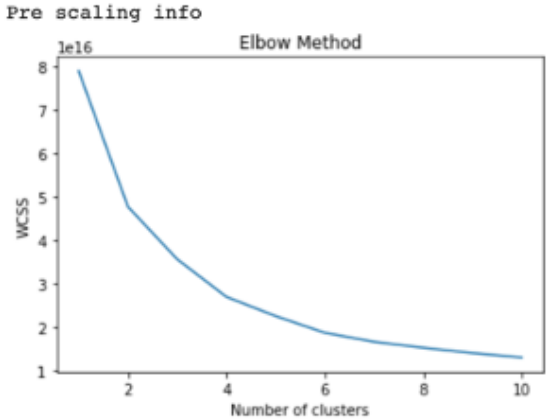

Fig. 10.    WCSS without Scaling

It was then rerun after scaling to ensure that there was not a significant difference between the two methods which would result in a further investigation or splitting into two different models. This is shown in Fig. 11.

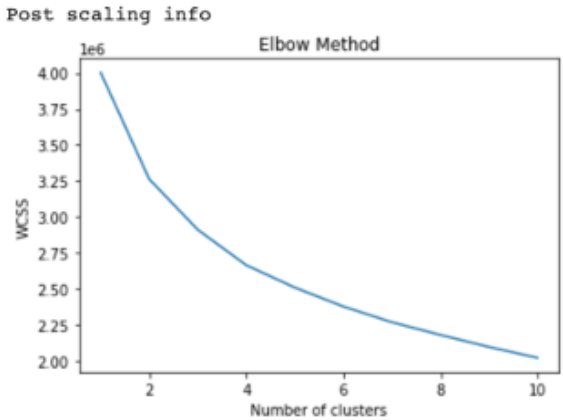

Fig. 11.    WCSS with Scaling

As we can see, the curve for pre-scaled information tapers off at 7 clusters. There is also a slight flattening of the curve at 7 clusters for post scaling information, which allowed us to maintain one model using 7 clusters.

K-Means was run using both SMOTE oversampling and RUS undersampling. K-Means was also run using min-max scaling as well as the standard scaling. These initial results are shown in Fig. 12.

As we can see, the results end up being poor, and while it outperforms randomly selecting one of the three options it falls in line with selecting the majority class each time. A further breakdown of the results is shown in Table I.

While the results were overall disappointing, they fit in line with preconceptions of K-Means on a dataset of this size in terms of the number of rows and number of attributes.

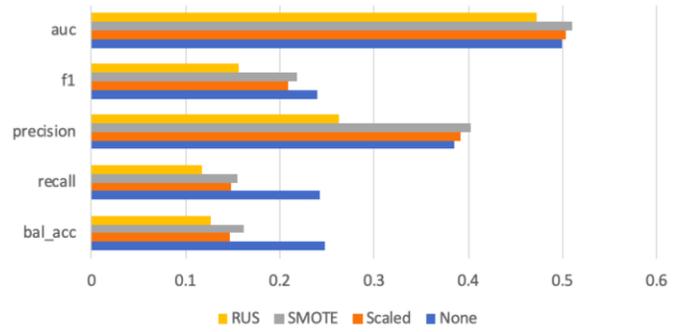

Fig. 12.    Initial K-Means Results

TABLE I.    K-MEANS 5-FOLD CV RESULTS

| Sampling and Scaling Experiments | | | | | |
|---|---|---|---|---|---|
| Model Type | Split | Recall | Precision | F1 | AUC |
| Baseline | Train | **0.2432** | **0.3870** | **0.2406** | **0.5001** |
| | Test | **0.2430** | **0.3855** | **0.2402** | **0.4999** |
| MM | Train | 0.1477 | 0.3924 | 0.2095 | 0.5030 |
| | Test | 0.1477 | 0.3920 | 0.2094 | 0.5029 |
| SMOTE + MM | Train | 0.1560 | 0.4042 | 0.2194 | 0.5106 |
| | Test | 0.1555 | 0.4032 | 0.2188 | 0.5104 |
| RUS + MM | Train | 0.1179 | 0.2624 | 0.1570 | 0.4730 |
| | Test | 0.1177 | 0.2621 | 0.1567 | 0.4729 |

## C. Support Vector Machine

Due to the large sample size, Stochastic Gradient Descent (SGD) classifier from Sklearn library was used with *hinge loss* in order to implement a regularized version of SVM model. This significantly reduced the time to run each model.

SVM was run through 5-fold validation in order to reduce the impact of outliers and extraneous attributes. The modeling was run with both SMOTE oversampling and RUS undersampling in order to test both results. Both standard and minmax scaling were used as well.

Every single scaling or sampling method was used individually then each combination of scaling and sampling method was used together. This allowed for the ability to test the impact of each. The results of these different combinations are shown in Table II.

TABLE II.    SUPPORT VECTOR MACHINE 5-FOLD CV RESULTS

| Sampling and Scaling Experiments | | | | | |
|---|---|---|---|---|---|
| Model Type | Split | Recall | Precision | F1 | AUC |
| Baseline | Train | 0.4711 | 0.5067 | 0.4220 | 0.5809 |
| | Test | 0.4712 | 0.5072 | 0.4220 | 0.5809 |
| RUS | Train | 0.4238 | 0.5205 | 0.3627 | 0.5688 |
| | Test | 0.4236 | 0.5186 | 0.3624 | 0.5687 |
| SMOTE  | *Train* | 0.4250 | 0.5549 | 0.3909 | 0.5881 |
|        | *Test*  | 0.4248 | 0.5547 | 0.3908 | 0.5879 |
| RUS + MM | *Train* | 0.5981 | 0.5629 | 0.5067 | 0.6839 |
|        | *Test*  | 0.5981 | 0.5551 | 0.5068 | 0.6840 |
| RUS + STD | *Train* | 0.5987 | 0.5663 | 0.5158 | 0.6878 |
|        | *Test*  | 0.5986 | 0.5647 | 0.5156 | 0.6878 |
| SMOTE + MM | *Train* | 0.5956 | 0.5639 | 0.5054 | 0.6849 |
|        | *Test*  | 0.5955 | 0.5664 | 0.5053 | 0.6849 |
| SMOTE + STD | *Train* | **0.6003** | **0.5689** | **0.5171** | **0.6879** |
|        | *Test*  | **0.6003** | **0.5687** | **0.5171** | **0.6879** |

As seen in Table II, the best result came from SMOTE oversampling and a standard scaling, showing a 0.09 increase in test F1 and a 0.1 increase in AUC score from the baseline model. When looking at each model, we also looked at the time it took for each to finish training. In this investigation, SMOTE methods took slightly longer than RUS, or no sampling methods, but the time taken increased only slightly, approximately 33%. This could be a potential downfall despite SMOTE having the best results, but it was not deemed significant enough to prevent us from picking the best results.

Given the increases across the board, scaling always outperformed unscaled methods, showing the impact that scaling has on SVM modeling methods. Looking at SMOTE alone, scaled models had a 0.12 and 0.1 increase in test F1 and AUC scores respectively. This increase shows the benefit of scaling a model with an abundance of varied attributes.

### D. Logistic Regression

Similar to the implementation of SVM, the Logistic Regression probabilistic model was also implemented with SGD training but using the *log loss* instead. Combinations of the Standard scaling method with SMOTE and RUS sampling were tried to assess the impact of preprocessing techniques on the performance when compared with the baseline LR model. While the results in Table IV suggest that SMOTE+STD has the best test F1 score of 60.38% among other combinations, it also takes the longest to train. As a result, RUS+STD, with a test F1 score of 60.28% was selected for further experiments in order to build an overall efficient model. The higher values of test scores compared to that of train points to the presence of underfitting in our algorithm, which motivated us to further investigate the parameter values to increase the performance.

Some important parameters to tune when it comes to LR is the type and strength of regularization to apply to the features. The parameter C is used for specifying the inverse of regularization strength in a generic LR model from sklearn. However, since the SGD classifier was used instead, it's alpha parameter was utilized to control this regularization strength. Elastic Net regularization with alpha of 0.00003 produced the best test F1 score of 60.49% compared to other combinations. These parameters also helped eliminate the underfitting issue.

In order to find which of the features have a larger impact on the prediction performance, coefficients of the LR model were analyzed and the top 20 were filtered, as shown in Fig. 13. The relation between some of the features and the target of grade is obvious. For example, having a longer loan term reduces your grade and puts you in a higher risk category, which is explained by the highly negative coefficient value. On the other hand, the large positive coefficient of the latest fico score suggests that the grade increases with this feature. However, a surprising discovery was the presence of a number of features that correspond to the purpose of the loan itself.

Taking out a loan for small business or debt consolidation purposes comes with a higher risk for the investors, which in turn negatively affects your grade. But, home improvement and other house related loans increases a borrower's grade and automatically puts them in a lower risk category. This can also be seen in the positive coefficient values for features such as if the borrower's homeownership is of the type mortgage and the number of mortgage accounts they in turn have.

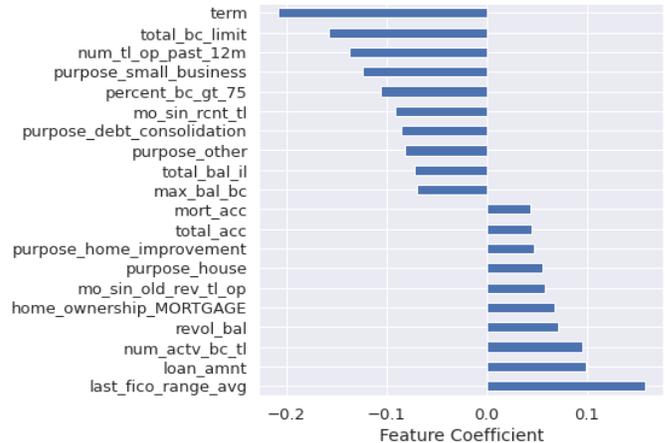

Fig. 13. Top 20 Features of Logistic Regression

In addition to analyzing these top coefficients, further attention was also placed in finding out which of the features have the least impact on the grade. A total of six such features were found with zero coefficients, as seen in Table III. Majority of these features correspond to a borrower's delinquency, charge-offs, and accounts that are past due, and it is surprising to find that they don't have an impact on grade.

TABLE III. ZERO COEFFICIENT FEATURES OF LOGISTIC REGRESSION

| ['mths_since_last_delinq', 'tot_cur_bal', 'collections_12_mths_ex_med', 'num_accts_ever_120_pd', 'num_tl_90g_dpd_24m', 'chargeoff_within_12_mths'] |
|---|

The performance metrics of an LR model with all the features were compared with those of a model with the top 20 features and another without the zero-coefficient features. As shown in Table IV, a test F1 score of 60.47% was achieved without the low-impact features, which is slightly lower than the best test F1 score of 60.49%. On the other hand, the results of just the top 20 features were not that impressive. Such a result can be expected with the LR model's mechanism of using actual coefficient values to rank the feature importances.

TABLE IV. LOGISTIC REGRESSION 5-FOLD CV RESULTS

| Sampling and Scaling Experiments | | | | | |
|---|---|---|---|---|---|
| Model Type | Split | Recall | Precision | F1 | AUC |
| Baseline | Train | 0.4717 | 0.5168 | 0.3883 | 0.5711 |
| | Test | 0.4717 | 0.5155 | 0.3883 | 0.5712 |
| STD | Train | 0.6293 | 0.5983 | 0.5969 | 0.6930 |
| | Test | 0.6292 | 0.5981 | 0.5968 | 0.6929 |
| SMOTE + STD | Train | 0.6160 | 0.6067 | 0.6036 | 0.7073 |
| | Test | 0.6162 | 0.6070 | 0.6038 | 0.7075 |
| RUS + STD | Train | 0.6137 | 0.6066 | 0.6026 | 0.7067 |
| | Test | 0.6140 | 0.6068 | 0.6028 | 0.7069 |
| Parameter Tuning Experiments with RUS + STD | | | | | |
| L2 + Alpha $3 * 10^{-5}$ | Train | 0.6147 | 0.6062 | 0.6011 | 0.7074 |
| | Test | 0.6147 | 0.6062 | 0.6011 | 0.7075 |
| Elastic Net + Alpha $3 * 10^{-5}$ | Train | **0.6139** | **0.6089** | **0.6052** | **0.7070** |
| | Test | **0.6136** | **0.6086** | **0.6049** | **0.7068** |
| Feature Selection Experiments with Elastic Net + Alpha 0.00003 | | | | | |
| Top 20 Features | Train | 0.5774 | 0.5672 | 0.5628 | 0.6759 |
| | Test | 0.5776 | 0.5673 | 0.5630 | 0.6760 |
| Without Zero-Coef Features | Train | 0.6132 | 0.6091 | 0.6049 | 0.7071 |
| | Test | 0.6130 | 0.6088 | 0.6047 | 0.7068 |

### E. Random Forest

The Random Forest classifier is an upgraded Decision Tree algorithm via the ensemble learning approach using bootstrap aggregation. The model is prone to overfitting issues, as seen in the results of scaling and sampling experiments in Table V. A combination of RUS+STD helped reduce overfitting while also increasing the test F1 score to 64.26%, and will be used for further experiments. Similar to the LR model, different parameter and feature combinations were tried out here too.

The number of trees in a RF model has a huge impact on the performance metrics and overfitting issues. The minimum samples required at a particular leaf node before a split is another parameter with similar characteristics. A combination of 120 trees along with 20 minimum leaf samples produces the best test F1 score of 64.61%, which is an improvement from the results before parameter tuning. Additionally, overfitting has also been reduced significantly, as shown in Table V.

As the next step in our analysis process, the top 20 features were filtered based on their importance as shown in Fig. 14. In contrast to the LR model which assigns positive or negative coefficients based on their impact on the target, RF outputs an array of importances regardless of their correlation with grade.

Similar to the LR model, some of the features in this top list are expected like annual income and total current balance. There are 8 features that appear in both the LR and RF lists such as term, loan amount, number of new accounts opened in the past 12 months, and so on. In addition to the latest fico score, the original fico value is also considered as important, which could help highlight the change in a borrower's spending patterns and/or paying-back ability.

Taking the loan for the credit card is the only purpose feature that appears in the list, in contrast to the LR model. An interesting find is the appearance of both the dti and dti-joint features, which suggest that having a joint-application and the debt-to-income ratio of both the borrowers have a major impact on grade prediction.

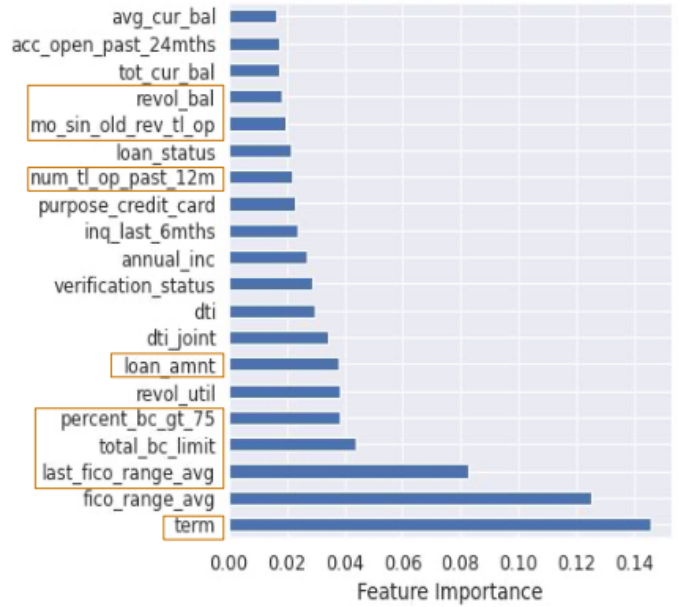

Fig. 14. Top 20 Features of Random Forest

Performance of the RF model with all the features together was compared with those of the top 20, the top 25, and the top 30 features. As shown in Table V, the results increase with the total number of features used, and there is a steady improvement. This shows there is a possibility of achieving the best test F1 score of 64.61% and test AUC score of 73.04% with a smaller feature subset or even get better scores with further tuning of the parameters.

TABLE V. RANDOM FOREST 5-FOLD CV RESULTS

| Sampling and Scaling Experiments | | | | | |
|---|---|---|---|---|---|
| Model Type | Split | Recall | Precision | F1 | AUC |
| Baseline | Train | 1.0000 | 1.0000 | 1.0000 | 1.0000 |
| | Test | 0.6535 | 0.6390 | 0.6415 | 0.7186 |
| STD | Train | 1.0000 | 1.0000 | 1.0000 | 1.0000 |
| | Test | 0.6536 | 0.6392 | 0.6416 | 0.7189 |
| RUS + STD | Train | 0.8977 | 0.9045 | 0.8981 | 0.9262 |
| | Test | 0.6368 | 0.6523 | 0.6426 | 0.7290 |

| Parameter Tuning Experiments with RUS + STD | | | | | |
|---|---|---|---|---|---|
| 10 Trees + Min 50 Leaf Samples | *Train* | 0.7807 | 0.7886 | 0.7827 | 0.8349 |
| | *Test* | 0.6381 | 0.6523 | 0.6437 | 0.7290 |
| 120 Trees + Min 20 Leaf Samples | *Train* | **0.7246** | **0.7348** | **0.7281** | **0.7927** |
| | *Test* | **0.6405** | **0.6544** | **0.6461** | **0.7304** |
| Feature Selection Experiments with 120 Trees + Min 20 Leaf Samples | | | | | |
| Top 20 Features | *Train* | 0.7075 | 0.7197 | 0.7118 | 0.7800 |
| | *Test* | 0.6342 | 0.6500 | 0.6405 | 0.7257 |
| Top 25 Features | *Train* | 0.7181 | 0.7301 | 0.7222 | 0.7880 |
| | *Test* | 0.6377 | 0.6537 | 0.6441 | 0.7285 |
| Top 30 Features | *Train* | 0.7217 | 0.7330 | 0.7256 | 0.7906 |
| | *Test* | 0.6388 | 0.6541 | 0.6449 | 0.7292 |

## VI. Discussion

Following the initial runs of each model, we contrasted the top two performing models, Logistic Regression and Random Forest, against each other on the unseen 20% test split of the original dataset. For the Logistic Regression, we took the RUS+STD preprocessing configuration with the parameters of Elastic Net regularization plus an alpha of 0.00003. With the Random Forest model, we applied RUS+STD preprocessing as well but with the parameters of 120 trees plus a minimum of 20 samples per leaf node. Both these models were trained on the entire 80% train split of the data.

As shown in Table VI, Random Forest performs better on each metric, with a F1 score of 65% compared to a F1 of 60% on the Logistic Regression model. Additionally, the AUC score of RF is also better with a value of 73.10% compared to that of LR, which is 70.60%. Assuming all else equal, it can be seen that the Random Forest is able to predict the grade for a random row better than Logistic Regression.

TABLE VI. Test Data Results

| Model Type | Recall | Precision | F1 | AUC |
|---|---|---|---|---|
| Logistic Regression | 0.6100 | 0.6100 | 0.6000 | 0.7060 |
| Random Forest | **0.6400** | **0.6600** | **0.6500** | **0.7310** |

For business purposes, however, it may not be better to predict on average, but instead, be able to predict extremes. For example, a lender may be more interested in highly risky applicants because they can impact their return on investment. Likewise, low-risk applicants can vastly improve return on investment as they are much less likely to default.

In order to investigate the breakdown of each model, we looked at a normalized confusion matrix of both the Linear Regression and Random Forest models. Fig. 15 shows both models' normalized confusion matrices, with a detailed per-class breakdown of accuracy highlighted in each row.

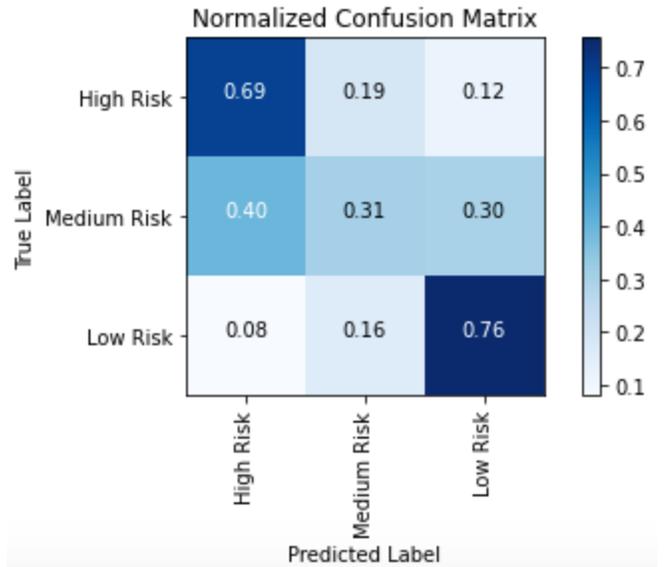

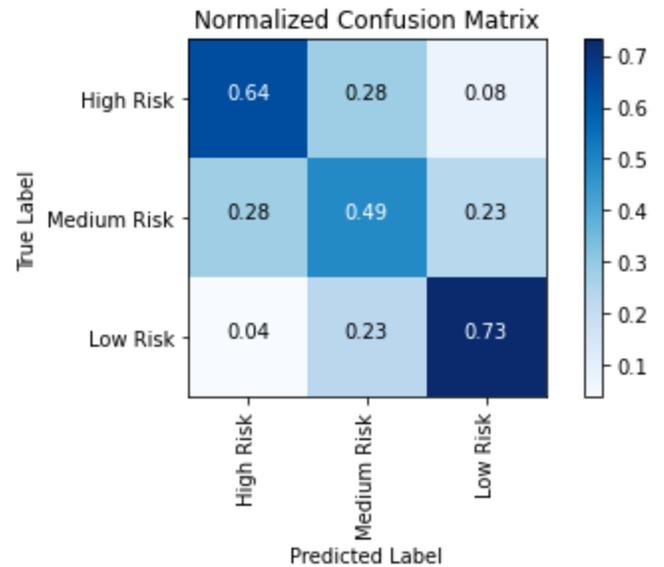

Fig. 15. Test Data Confusion Matrix: Top - LR, Bottom - RF

As we can see in the confusion matrices, the polar points of high risk/high risk and low risk/low risk are better in Linear Regression than Random Forest. Therefore, if looking to predict the extreme cases which may have a larger impact on business viability and return on investment, it would be best to use Linear Regression. However, the advantage is slight and therefore may not be worth it, given the pullbacks in performance on the medium-risk category.

Looking at each of these models closer gives an important view of what may be important to each company. While we are unable to predict what would be more beneficial given the needs and desires of financial institutions, looking at each model in further detail will allow us to provide different options depending on the needs and desires of said company.

## VII. Conclusion And Future Work

Beyond the initial results and predictive analysis of Logistic Regression on the extreme ends of the risk spectrum and Random Forest on a general prediction, diving deeper into the features that influenced each result leads to interesting conclusions.

Looking at the top results for both Logistic Regression and Random Forest, multiple different purposes appear. Looking at the Logistic Regression graph, the purpose of small business loans and credit consolidation has a highly negative correlation whereas the purpose of home improvement has a highly positive correlation. This helps reaffirm the findings that Carlos Serrano-Cinca, Begoña Gutiérrez-Nieto, and Luz López-Palacios found in [11]. While there are numerous personal factors that come into play, the purpose of a loan also greatly influences the grade one would receive. This falls in line with the likelihood of each purpose for default found by Serrano-Cinca et. al. There are also a number of different attributes related to credit utilization rather than total debt or total credit. This can be used to show that the percentage of credit used plays outsized importance relative to income.

Looking at future testing, it would be beneficial to look at each of these two subsets in their entirety and determine whether purpose alone or credit use alone can determine the grade someone will receive. Looking at individual subsets such as these can be beneficial.

The majority of the modeling methods can be improved via ensemble learning. Thus, investigating the benefits of combining Logistic Regression with Random Forest techniques, and the impact it could have on the predictive nature of the models, is a good scope for future work. Further testing with different sampling methods and scaling methods can also be investigated, though their results may not be as extreme as leveraging the benefits from ensemble testing.

Finally, another potential future work would be removing the Fico Score references in their entirety to see if there is a significant dropoff in the overall performance of the models. When looking at secondary methods of credit evaluation, it could be beneficial to eliminate the primary method to see if it is playing an outsized weight on predictive analysis. Looking at the dropoff after removing the Fico Score features would be useful in this instance.